# Controllable Interlocking from Irregularity in Two-Phase Composites

*Chelsea Fox[1], Kyrillos Bastawros[1], Tommaso Magrini[2]\* and Chiara Daraio[1]\**

[1] Division of Engineering and Applied Science, California Institute of Technology, Pasadena, CA 91125, USA

[2] Department of Mechanical Engineering, Eindhoven University of Technology, 5600MB Eindhoven, The Netherlands

\* Email: t.magrini@tue.nl, daraio@caltech.edu

**Abstract**

Natural materials often feature a combination of soft and stiff phases, arranged to achieve excellent mechanical properties, such as high strength and toughness. Many natural materials have even independently evolved to have similar structures to obtain these properties. For example, interlocking structures are observed in many strong and tough natural materials, across a wide range of length scales. Inspired by these materials, we present a class of two-phase composites with controllable interlocking. The composites feature tessellations of stiff particles connected by a soft matrix and we control the degree of interlocking through irregularity of particle size, geometry and arrangement. We generate the composites through stochastic network growth, using an algorithm which connects a hexagonal grid of nodes according to a coordination number, defined as the average number of connections per node. The generated network forms the soft matrix phase of the composites, while the areas enclosed by the network form the stiff reinforcing particles. At low coordination, composites feature highly polydisperse particles with irregular geometries, which are arranged non-periodically. In response to loading, these particles interlock with each other and primarily rotate and deform to accommodate non-uniform kinematic constraints from adjacent particles. In contrast, higher coordination composites feature more monodisperse particles with uniform geometries, which collectively slide. We then show how to control the degree of interlocking as a function of coordination number alone, demonstrating how irregularity facilitates controllability.

**Introduction**

Nature offers an abundance of materials with excellent mechanical properties, including high stiffness[1–6], high strength[2–4,7–11], high toughness[1,4,5,10,12–15] and good energy absorption[16–18]. These materials are often composed of stiff and soft phases, arranged to optimize mechanical performance. Many biological materials have even independently evolved to have similar structures across a wide range of length scales[19]. For example, interlocking structures that provide excellent mechanical performances can be observed in many different biological materials, providing high strength, ductility and toughness[19]. In nacre, interlocking occurs as a result of rough, wavy tablets, which jam as they slide past one another[1,20,21], while in turtle carapaces and cranial bone, interlocking is seen in zigzag bone interfaces, which engage like puzzle pieces for improved bending strength and toughness[11,14,22]. Interlocking is even seen in stomatopod dactyl clubs, which feature bouligand and herringbone structures that deflect cracks with out-of-plane interlocked layer arrangement[23,24]. However, mimicking all of these advantageous biological structures for use in bioinspired engineering materials remains a complex design and fabrication challenge.

Here we propose the use of irregularity to generate bioinspired interlocking materials, and we present a class of two-phase composites composed of tessellations of stiff reinforcing particles connected by a soft matrix. Going one step beyond the biological materials, we show that our materials offer control over the degree of interlocking, defined as the kinematic constraints provided by neighboring particles[25,26], through control of particle size, geometry and arrangement. We generate the materials using a virtual growth algorithm (VGA), which mimics the growth of stochastic structures observed in nature by assembling simple building blocks into a network according to connectivity rules[27–31]. To further increase the design space, we present a VGA on a hexagonal grid (hexa-VGA), offering up to 6-sided connectivity. The hexa-VGA begins with a set of nodes on the hexagonal grid and randomly assigns connections from each node until the entire grid is filled, forming a network. We define the average number of connections per node as the coordination number[32,33], which we use as an input parameter to the hexa-VGA. The generated hexa-VGA network then forms the matrix phase of our composite materials, while the reinforcing particles are formed by the areas enclosed by the matrix.

As a function of coordination, and the resulting particle size, geometry and arrangement, particles interlock to varying degrees in response to the kinematic constraints provided by neighboring particles. This mechanical performance is reminiscent of not only interlocking biological materials, but also of interlocking engineering materials, which have been previously shown to provide tunable bending stiffness[34–37], enhanced load bearing capacity[36,38–40], and improved toughness[37,41–43]. However, all of the previous studies on interlocking are all limited by the periodic nature of the interlocking elements, in both the biological materials and the engineering materials.

In contrast, our irregular elements (particles) offer a wider design space, with control over the degree of interlocking in response to loading, as a function of coordination number.

**Results and Discussion**

*Sample Design*

To design and generate our samples, we use a hexagonal virtual growth algorithm (hexa-VGA), which stochastically grows a network from a set of nodes on a hexagonal grid. The hexa-VGA is defined by a coordination number, which is the average number of connections per node (Figure 1a). We can therefore define a set of 63 geometrically or rotationally unique hexagonal tiles, which form from the network on the grid (Figure 1b,c). To make our polymer composite materials, we additively manufacture the generated network as the soft matrix phase, while the areas enclosed by the network form the stiff reinforcing phase. Further details on the generation and fabrication methods may be found in the Methods section.

To span the available design space, we generate samples with coordinations of 2X, 2.5X, 3X, 3.5X, 4X, 5X, and 6X, which are composed of primarily either 100% the coordination tile type (for integer coordinations) or 50% of the tile type above and 50% of the tile type below (for non-integer coordinations) (Figure 1e). The particle size is inversely related to the coordination number, with coordination 2X below the percolation threshold for equilateral triangular networks[44], resulting in the largest particles (Figure 1f). The smallest particle size possible is that of a triangle formed by adjacent lines (Figure 1c).

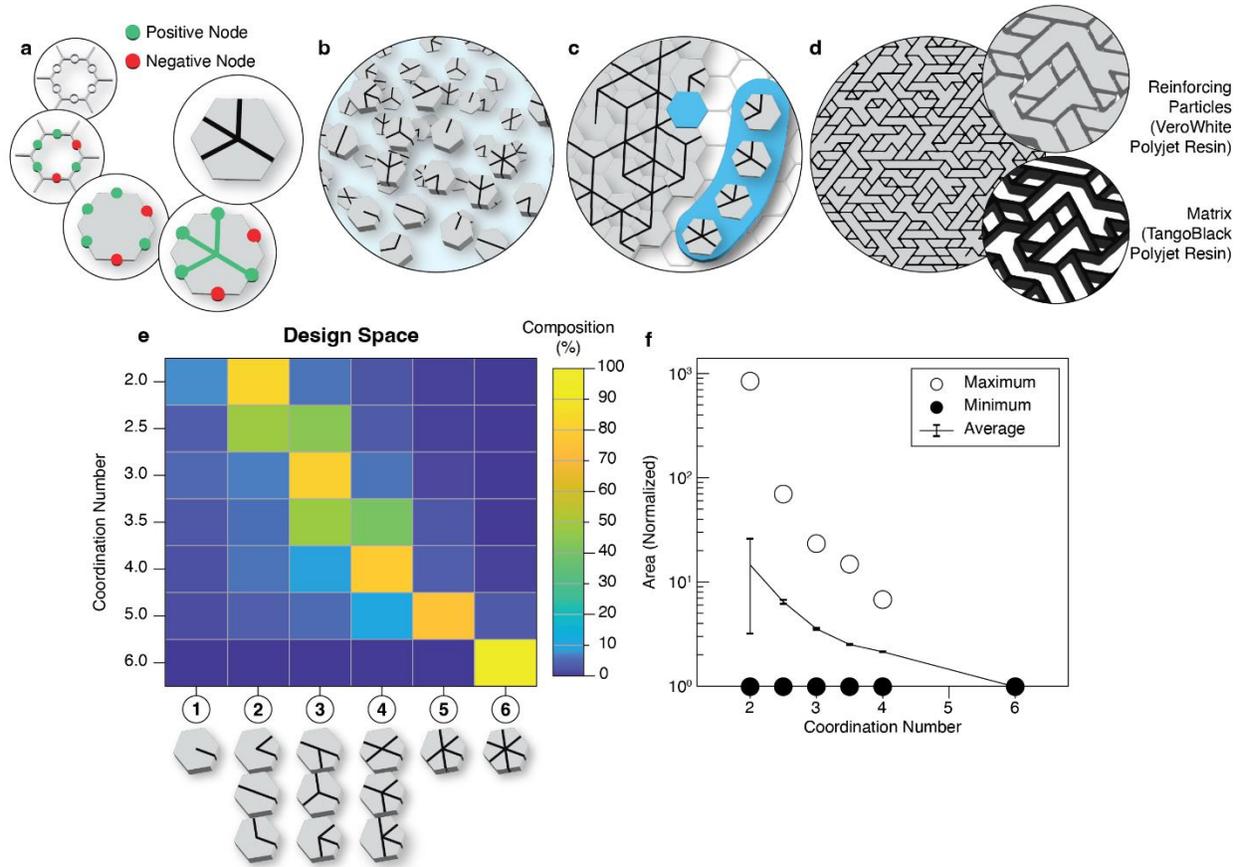

**Figure 1.** Sample generation and characterization of design space. (a) Node identification with positive (green) and negative (red) nodes. (b) Assorted hexagonal tiles. (c) Hexagonal tile connectivity. (d) Composite material generation with reinforcing particles and matrix. (e) Hexa-VGA sample compositions as a function of coordination number. (f) Triangle-normalized particle size distributions as a function of coordination number.

*Mechanical Characterization - Cylindrical Contact Loading*

Samples are loaded in compression with a cylindrical contact to understand how the structure responds to localized load at displacements up to 3 mm (Figure 2a-f). We test three different samples for each coordination (2X, 2.5X, 3X, 3.5X, 4X) and observe that at the lowest coordination, the material behaves similarly to a bulk material, as it falls below the percolation threshold for an equilateral triangular network[44] and is primarily composed of a few large particles (Figure 2a). As coordination number increases, stiffness and strength decrease non-linearly, as a result of both structure and material properties, since the volume fraction of the reinforcing phase decreases with increasing coordination number (Figure 2b-f).

Given the hexagonal nature of the hexa-VGA used to generate the samples, we examine the effect of orientation by testing samples at 0° (horizontal, blue) and 90° (vertical, red), such that the underlying hexagonal grid is aligned along the widest hexagon direction and the narrowest hexagon direction. At lower coordinations, there is significant anisotropy, with the vertical orientation displaying greater stiffness and strength (Figure 2a,b). This trend decreases as coordination number increases (Figure 2c-e), until we reach the periodic 6X case, where the vertical orientation remerges as being stiffer (Figure 2f, Figure S2). This anisotropy effect is likely the result of the matrix alignment with respect to the direction of loading, where horizontal alignment allows for greater deformability in the lateral x-direction when loaded from the normal y-direction.

We then use 2D digital image correlation (DIC) to track the sample deformation up to 1 mm cylindrical contact displacement (indicated by the gray line in the force-displacement plots). Across the various samples (Figure 2g-l, Figure S1), we observe varying amounts of strain surface area (structural engagement) as a function of coordination number, with intermediate coordinations displaying the largest region of both $\varepsilon_x$ (Figure 2m-r, Figure S1) and $\varepsilon_y$ strain (Figure 2s-x, Figure S1).

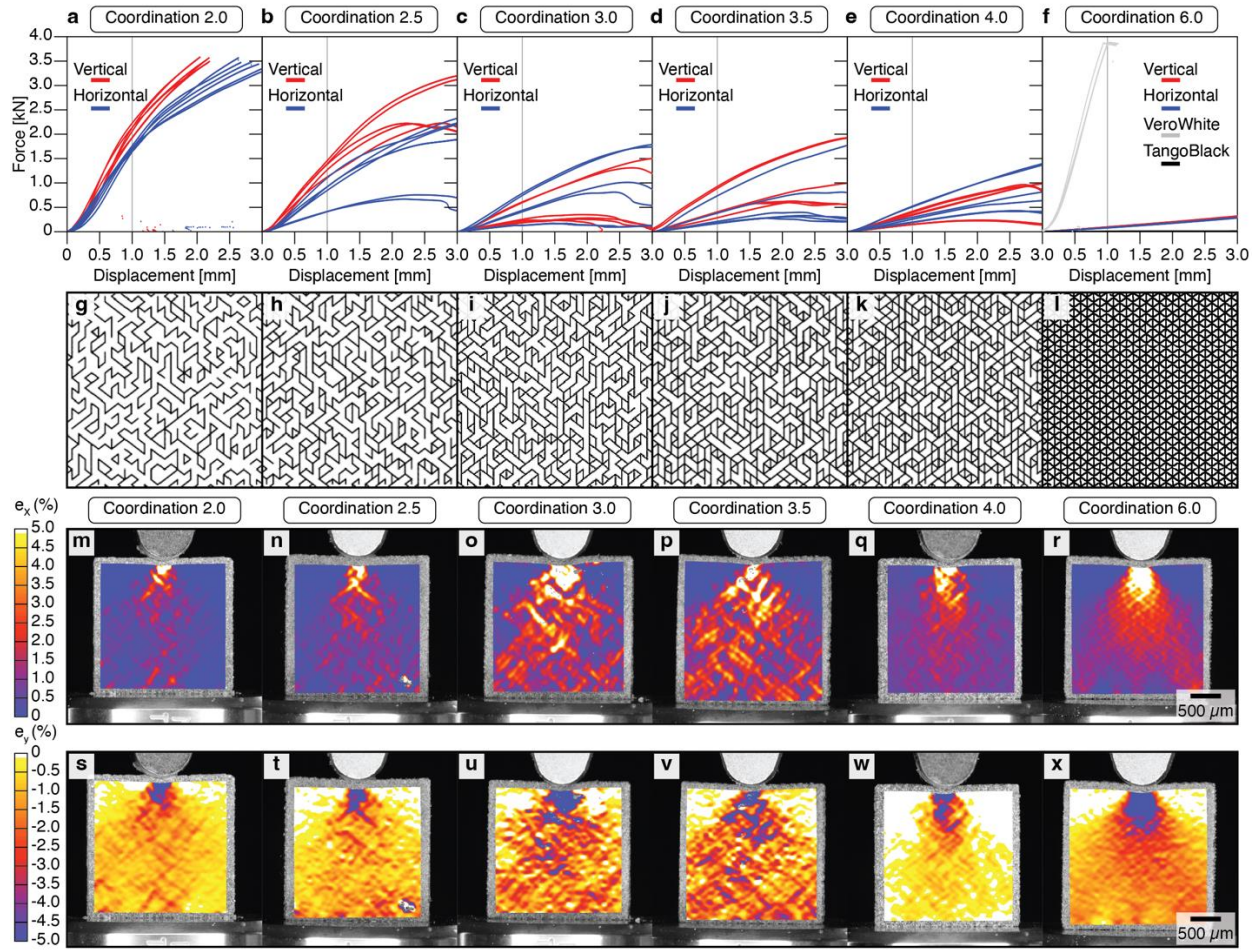

**Figure 2.** Cylindrical contact loading characterization. (a-f) Force-displacement plots for coordinations 2X, 2.5X, 3X, 3.5X, 4X and 6X, respectively. Red refers to vertical matrix orientation, blue refers to horizontal matrix orientation, gray line highlights 1 mm displacement. (g-l) Selected samples with vertical orientation. (m-r) 2D DIC maps of $\varepsilon_x$ strain for corresponding selected samples. (s-x) 2D DIC maps of $\varepsilon_y$ strain for corresponding selected samples.

*Mechanical Characterization - Matrix Response*

To understand the reason for the non-linear trend in the structural engagement, we first characterize the matrix response as a function of coordination number. The strain surface maps can be masked to obtain the strain maps of exclusively the soft matrix (Figure a-c). After normalizing for the volume fraction of matrix in each sample, we observe that matrix engagement (defined as non-zero matrix strain across the sample surface area) follows the same trend as the overall structure engagement and varies with coordination number. The lowest values occurring at 6X, followed by 2X, and the greatest values occurring around 3X to 3.5X, once we reach a cylindrical contact displacement greater than 0.33 mm, for both $\varepsilon_x$ and $\varepsilon_y$ strain (Figure 2d,e). This peak of matrix engagement is likely the result of particle geometries and arrangement, whose irregularity leads a series of kinematic constraints as particles engage with neighboring particles, distributing strain over a large amount of the sample.

To further quantify how the matrix distributes strain across the structure, we also measure the average strain across the sample depth. At 1 mm cylindrical contact displacement, we first convert the strain maps to grayscale values in MATLAB (MathWorks, USA) and then collapse the strain maps to a vertical line, where darker values indicate greater amounts of strain (Figure 2f-k). We observe that intermediate coordinations also display the greatest depth of matrix engagement (Figure 2l), with maxima around 3X and 3.5X (Figure 2m). This again indicates that these intermediate coordinations engage the largest amount of the composite structure in response to the cylindrical contact loading.

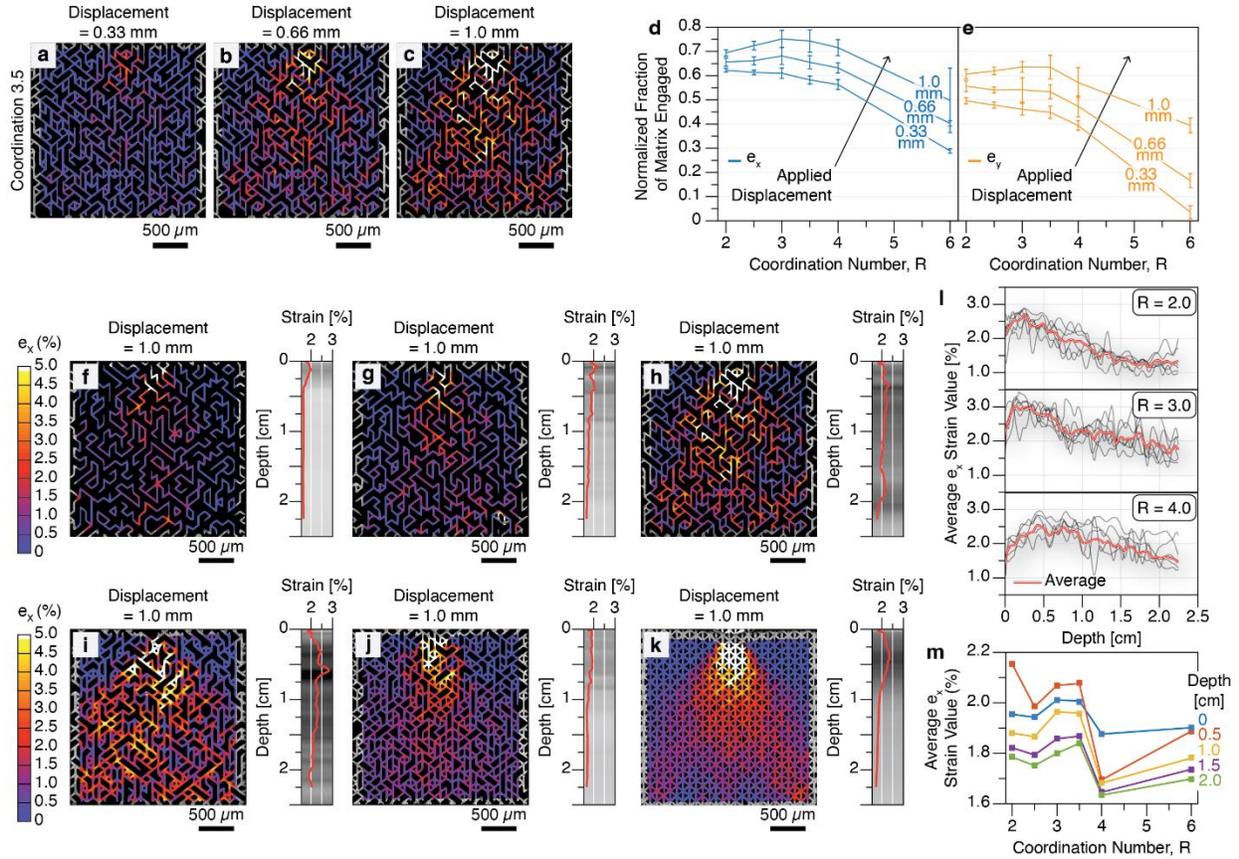

**Figure 3.** Matrix response characterization. (a-c) Example 3.5X matrix $\varepsilon_x$ strain with increasing cylindrical contact displacement. (d) Volume normalized fraction of matrix engaged as a function of coordination number and cylindrical contact displacement for $\varepsilon_x$ strain. (e) Volume normalized fraction of matrix engaged as a function of coordination number and cylindrical contact displacement for $\varepsilon_y$ strain. (f-k) Example 2X, 2.5X, 3X, 3.5X, 4X and 6X matrix $\varepsilon_x$ strains at 1 mm cylindrical contact displacement with corresponding horizontally averaged $\varepsilon_x$ strain values (grayscale bars with line plot average). (l) Average $\varepsilon_x$ strain value as a function of depth for all 2X, 3X and 4X coordinations at 1 mm cylindrical contact displacement. (m) Average $\varepsilon_x$ strain value as a function of coordination number for all 2X, 3X and 4X coordinations at various sample depths.

*Mechanical Characterization – Particle Response*

We then examine the particle response to further understand the relationship between coordination number and how the structure accommodates the applied loading. In this study, we focused on the low-strain regime, and therefore it is important to note the tough matrix-particle interface does not fracture during loading[45] and particles remain adhered to their surrounding matrix phase throughout our analysis.

The 2X samples are primarily composed of a few large particles, given that they are below the percolation threshold, and behave similarly to a bulk material. As the coordination increases and crosses the percolation threshold, particle number increases and particle size decreases, although lower coordinations (3X) still feature highly polydisperse particles with irregular geometries which are often concave (Figure 4a). Under cylindrical contact loading, these low-coordination samples deform as individual particles uniquely translate, rotate and deform to accommodate the non-uniform kinematic constraints provided by neighboring particles. We define the degree to which particles are kinematically constrained by neighboring particles as interlocking[25,38] and the lower coordinations display the greatest amount of interlocking (Figure 4a, SI Video 1). To quantify the interlocking behavior, individual particle path vectors can be tracked using FIJI TrackMate[46], and we can then use these vectors to observe how the particles move relative to one another (Figure 4b). Greater interlocking results in a particle vector that is more dissimilar to neighboring particle vectors, resulting in a wide distribution of vector angles in a local region (Figure 4c). As the coordination increases further to 4X, particle size continues to decrease, and particles become more uniform in both shape and size (Figure 4d). In addition to particle-to-particle interlocking, this uniformity results in the activation of mechanisms of collective particle sliding, due to the reduced neighboring particle kinematic constraints in local regions (SI Video 2). This mixed mode behavior is reminiscent of nacre, although the interlocking and collective sliding of nacre's tablets is a sequential deformation response resulting from monodisperse tablets[1,9,20,47], rather than a simultaneous tradeoff resulting from polydispersity and irregularity. The mixed mode behavior results in a narrower distribution of particle vector angles, as the collectively sliding particles have more similar angles in a local region (Figure 4e,f). At 6X coordination, all particles collectively slide, as all particles are convex and periodically arranged, with uniform shape and size (Figure 4g,h, SI Video 3), and the distribution of particle vector angles becomes much narrower in a local region (Figure 4i).

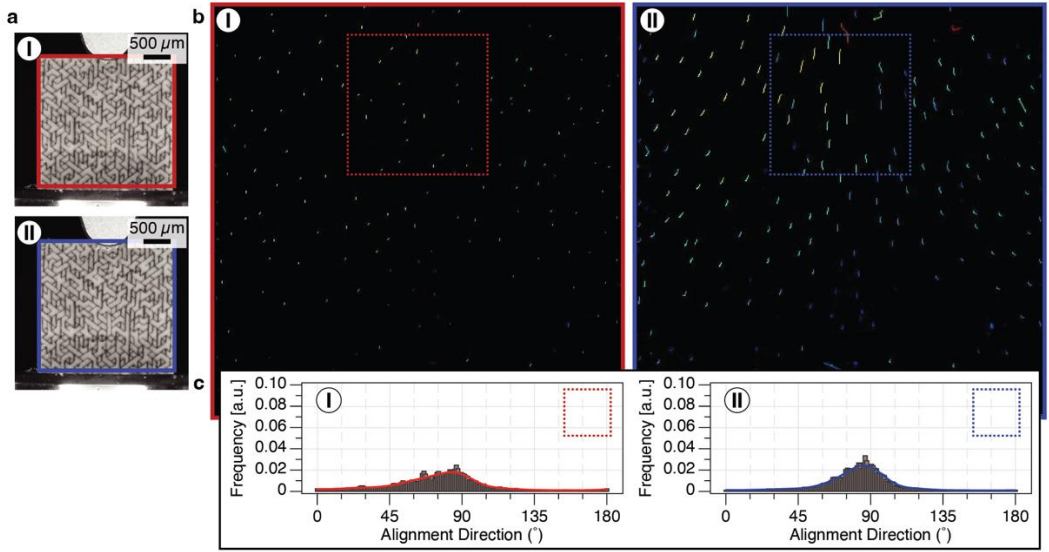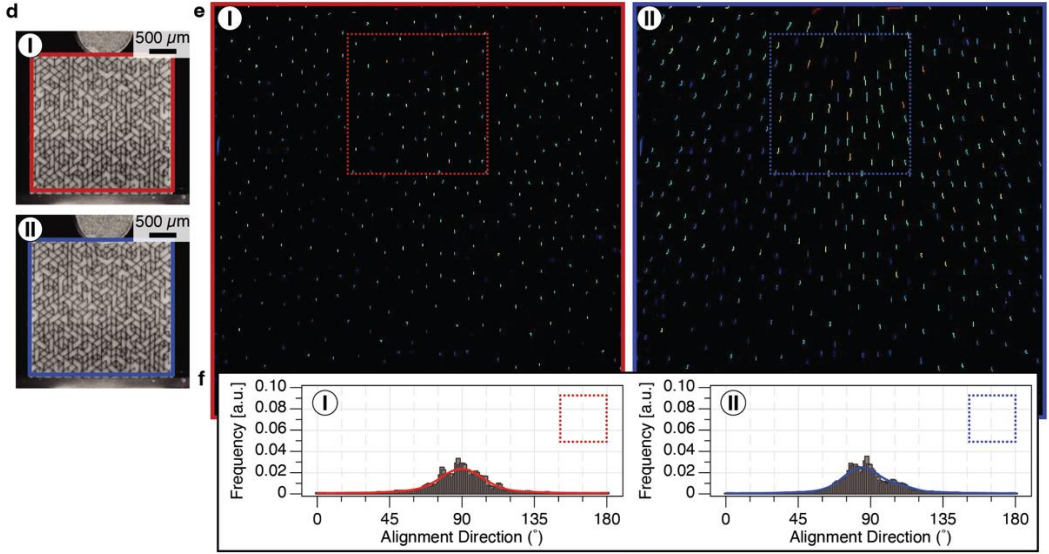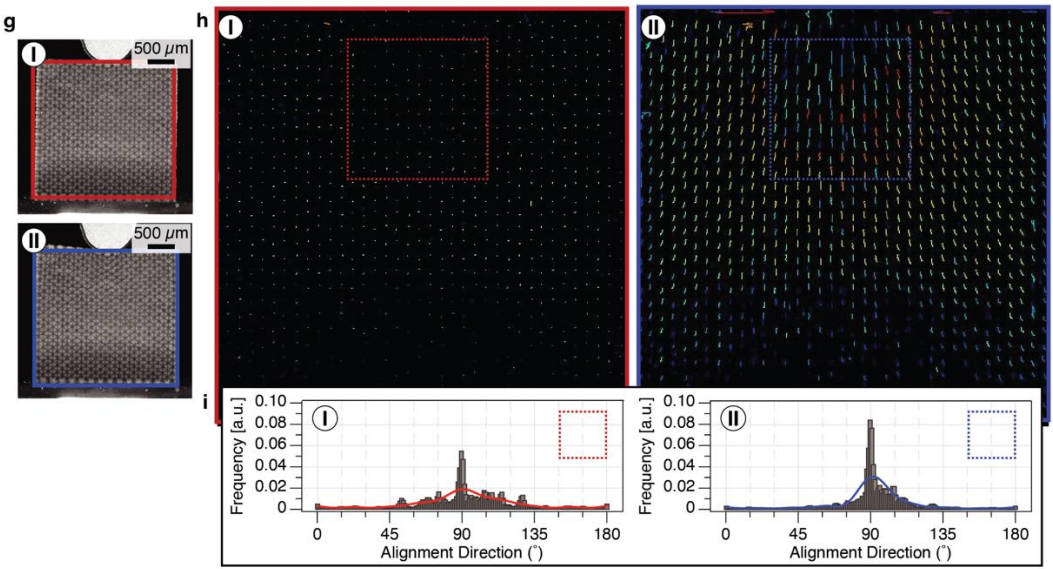

**Figure 4.** Particle response characterization. (a) Example 3X particle image at 0.5 mm (I) and 1 mm (II) cylindrical contact displacement. (b) Corresponding particle vector map at 0.5 mm (I) and at 1 mm (II) displacement. (c) Histogram of vector map line angle frequencies at 0.5 mm (I) and 1 mm (II) displacement. (d) Example 4X particle image at 0.5 mm (I) and 1 mm (II) displacement. (e) Corresponding particle vector map at 0.5 mm (I) and at 1 mm (II) displacement. (f) Histogram of vector map line angle frequencies at 0.5 mm (I) and 1 mm (II) displacement. (g) Example 6X particle image at 0.5 mm (I) and 1 mm (II) displacement. (h) Corresponding particle vector map at 0.5 mm (I) and at 1 mm (II) displacement. (i) Histogram of vector map line angle frequencies at 0.5 mm (I) and 1 mm (II) displacement.

*Mechanical Characterization - Statistical Analysis*

To quantify the transition from interlocking to collective sliding behavior in our materials, we examine the statistics behind the particle and matrix arrangement. To reduce interlocking and achieve collective sliding behavior, the matrix must be arranged in continuous straight lines to form planes along which particles can slide. Given an initial matrix orientation on a hexagonal tile, we can therefore determine which subsequent tiles allow the straight line of matrix to continue (Figure 5a), and which divert it (Figure 5b). With rotational symmetry, regardless of the initial matrix orientation, the tile distributions for continuing or diverting remains the same, resulting in a continuous or discontinuous line of matrix (Figure 5c). Given the input parameter of coordination number, which tells us which tile types we have available, we can then calculate the probability of the matrix continuing to determine the average length of matrix lines. We use Bayes theorem,

$$P(R \cap slip\ plane) = \left[\sum_1^i P(slip\ plane | tile\ type_i) * P(tile\ type_i)\right]^j \qquad (1)$$

where $i$ is the number of tile types, and $j$ is the number of tiles. By placing a threshold at a percent less than 1% likelihood, we can plot the expected length of straight lines of matrix for each coordination type and we can see that the probability increases non-linearly with coordination number (Figure 5d). Low coordinations have statistically shorter lines of continuously aligned matrix, resulting in complex geometries that interlock, while higher coordinations have statistically longer lines, with 6X showing only continuous lines, resulting in collective sliding behavior (Figure 5d).

To further understand what gives rise to interlocking behavior and how particles engage with neighboring particles, we also quantify the average number of neighboring particles per particle. Using the Euler characteristic[48], we determine the number of particles and the number of edges (which correspond to neighbors). The Euler characteristic is defined for a 2D graph as

$$V - E + F = 1, \qquad (2)$$

where $V$ is the number of tiles greater than 2X, $E$ is the number of edges, defined as

$$E = \frac{R_c V}{2}, \qquad (3)$$

where $R_c$ is the corrected coordination number, found by removing any 2X tiles, which only contribute to the length of the edges but not to the number of edges, and $F$ is the number of particles. $F$ can therefore be rewritten as

$$F = 1 - V(1 - \frac{R_c}{2}). \qquad (4)$$

We then determine the average number of neighboring particles per particle, $N$ (Figure 5e):

$$N = \frac{E}{F}. \qquad (5)$$

From the Euler characteristic (Equation 2), the maximum average number of neighbors per particle cannot exceed six, and as the corrected coordination number increases above 3X, the maximum number of neighbors per particle decreases, until we reach a limit of three neighbors at 6X (Figure 5f). This follows the trend we observed where the maximum amount of structural engagement occurs at intermediate coordinations. At these coordinations, we have both interlocking particles engaging with a nearly maximum number of neighbors, as well as collectively sliding particles, which easily move in large groups. At these intermediate coordinations, the average particle size also maintains a concave shape, which allows for more kinematic constraints (greater interlocking), formed by more than three triangles (Figure 5g). These kinematic constraints then allow the particles to easily distribute the loading to their neighboring particles across the structure.

To find the upper bounds of the design space, we plot together maximum continuous tilings and corresponding neighbors per particle, and we include our tested samples as well as additional samples which were only statistically analyzed (Figure 5h). To achieve the greatest degree of interlocking, lower coordinations are desirable, while the lowest degree of interlocking is achieved at 6X, when all particles collectively slide along continuous matrix lines. However, as previously discussed, the greatest amount of structural engagement occurs at intermediate coordinations, when there is a tradeoff between the simultaneous activation of the interlocking and collective sliding mechanisms. It should be noted that the upper limit defined here is for the homogenous case where samples are formed by nearly 100% of their coordination number tile type for integer coordinations or 50% above and 50% below for non-integer coordinations (Figure 1e). It is therefore possible to increase the continuous tiling length by forming the same coordination number with other tile compositions that have greater numbers of high coordination tiles, although these may not be feasible to generate using the hexa-VGA.

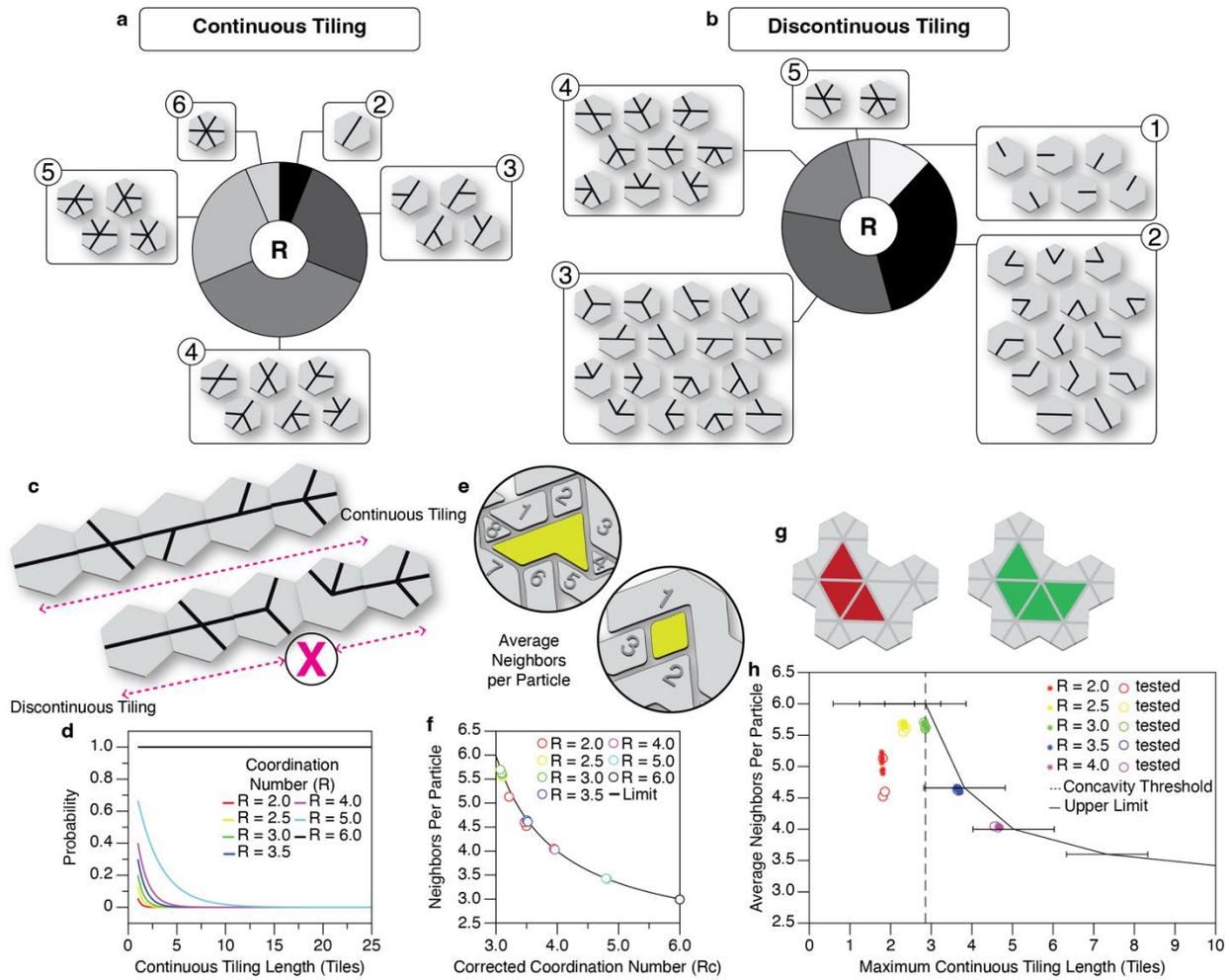

**Figure 5.** Statistical characterization of matrix and particles. (a) Continuous tiles for 60° example matrix plane according to coordination number. (b) Discontinuous tiles for 60° example matrix plane according to coordination number. (c) Example of continuous and discontinuous tiling plane. (d) Probability of continuous tiling as a function of number of tiles for various coordination numbers. (e) Examples of counts of neighboring particles per particle. (f) Neighboring particles per particle as a function of corrected coordination number, solid black line denotes upper limit. (g) Example of convex (red) particle and concave (green) and geometries. (h) Average neighboring particles per particle as a function of maximum continuous tiling, dashed black line denotes concavity threshold, solid black line denoted upper limit.

**Conclusions**

We present a class of two-phase composites composed of tessellations of stiff particles connected by a soft matrix. Drawing inspiration from the excellent mechanical performance of interlocking structures observed in many natural materials, our composites feature particle interlocking in response to loading. Going one step beyond the biological materials, we control the degree of interlocking using irregularity of particle size, geometry and arrangement. We generate the composites through a hexa-VGA, which stochastically connects a network of nodes on a hexagonal grid according to coordination number. We then use the generated hexa-VGA network as the matrix phase, while the reinforcing particles are formed by the areas enclosed by the matrix. Lower coordinations feature highly polydisperse particles, which interlock as a result of their irregular geometries and non-periodic tessellations. Higher coordinations feature more monodisperse particles, which collectively slide as a result of their more uniform geometries and tessellations. Finally, we show and statistically quantify how to control the tradeoff between these interlocking and sliding mechanisms, with activation of a particular mechanism and the amount of structural engagement controlled by coordination number alone.

**Methods**

*Sample Generation and Fabrication*

The hexa-VGA generates (irregular) networks by beginning with an equilateral triangular network placed on a hexagonal grid. Lists of unique lines in the network are first defined by their endpoints and then initiated with a status of "neither." An arbitrary endpoint (a node) is randomly selected, from which, X (out of 6) lines are assigned the status "positive," while the rest are given the status "negative". This number X is defined as the coordination number, and we define it on a set of tiles formed by the hexagonal grid. Once a line has been assigned a status, that status cannot change. After all lines beginning at an endpoint have been assigned a status, that endpoint is removed from the set of "free" endpoints. If any of the neighboring 6 endpoints are free endpoints, one of them is chosen to have its lines assigned a status. If none of the neighboring endpoints are free endpoints, a random free endpoint is selected, and this process continues until no free endpoints remain and all lines have been assigned a status, resulting in a network of interconnected nodes.

To fabricate our composite samples from the hexa-VGA output network, we first use the network information to create an STL file for the soft matrix, which is made from TangoBlack Polyjet Resin, and then take its negative to form the stiff reinforcing particles, which is made from VeroWhite Polyjet Resin. The mechanical properties of the two phases fall within those reported in literature[49–51]. Due to printer resolution constraints, we choose a matrix width of 100 μm and a minimum particle width of 1 mm.

The hexa-VGA code used to generate the samples is available from the authors upon reasonable request.

***Cylindrical Contact Compression Testing***

We use an Instron E3000 with a 5 kN load cell (Instron, USA) to apply compression loading. Samples of 2.5 cm by 2.5 cm by 1 cm are loaded with a cylindrical contact of 1 cm diameter. Testing is conducted for displacements up to 3 mm, which is just prior to fracture. Three different samples are tested for each coordination to ensure consistency across samples.

*Digital Image Correlation*

To conduct 2D digital image correlation (DIC) on the samples, we apply a layer of matte white paint and then matte black speckles with a diameter of 0.1-0.3 mm to the front face of the samples. During loading, we use a Nikon D750 DSLR camera (Nikon, Japan) with a 120mm lens to take images at a rate of 1 frame per second. We use VIC 2D (Correlated Solutions, USA) to conduct the DIC analysis, using a step size of 2 and a subset size of 29, and obtain the Lagrangian strain fields in the x-direction and y-direction.

## Acknowledgements

The authors acknowledge MURI ARO W911NF-22-2-0109 for the financial support. The authors acknowledge P. Arakelian for experimental assistance.

## Author Contributions

Conceptualization, C.F. and T.M.; Methodology, C.F. and T.M.; Software, C.F., K.B., and T.M.; Formal Analysis, C.F. and T.M.; Writing – Original Draft, C.F., K.B., and T.M.; Writing – Review & Editing, C.F., K.B., T.M., and C.D.; Visualization, C.F. and T.M.; Supervision, T.M. and C.D.; Funding Acquisition, C.D.

## Declaration of Interests

The authors declare no competing interests.

**References**

1. Barthelat, F., Tang, H., Zavattieri, P.D., Li, C.M., and Espinosa, H.D. (2007). On the mechanics of mother-of-pearl: A key feature in the material hierarchical structure. J Mech Phys Solids *55*, 306–337. https://doi.org/10.1016/J.JMPS.2006.07.007.

2. Oftadeh, R., Perez-Viloria, M., Villa-Camacho, J.C., Vaziri, A., and Nazarian, A. (2015). Biomechanics and Mechanobiology of Trabecular Bone: A Review. J Biomech Eng *137*, 0108021. https://doi.org/10.1115/1.4029176.

3. Hart, N.H., Nimphius, S., Rantalainen, T., Ireland, A., Siafarikas, A., and Newton, R.U. (2017). Mechanical basis of bone strength: influence of bone material, bone structure and muscle action. J Musculoskelet Neuronal Interact *17*, 114.

4. Magrini, T., Libanori, R., Kan, A., and Studart, A.R. (2021). Complex Materials: The Tough Life of Bone. Revista Brasileira de Ensino de Fisica *43*, 1–17. https://doi.org/10.1590/1806-9126-RBEF-2020-0438.

5. Launey, M.E., Buehler, M.J., and Ritchie, R.O. (2010). On the mechanistic origins of toughness in bone. Annu Rev Mater Res *40*, 25–53. https://doi.org/10.1146/ANNUREV-MATSCI-070909-104427/CITE/REFWORKS.

6. Lin, E., Li, Y., Weaver, J.C., Ortiz, C., and Boyce, M.C. (2014). Tunability and enhancement of mechanical behavior with additively manufactured bio-inspired hierarchical suture interfaces. J Mater Res *29*, 1867–1875. https://doi.org/10.1557/JMR.2014.175.

7. Ritchie, R.O. (2011). The conflicts between strength and toughness. Nature Materials 2011 10:11 *10*, 817–822. https://doi.org/10.1038/nmat3115.

8. Mirkhalaf, M., Dastjerdi, A.K., and Barthelat, F. (2014). Overcoming the brittleness of glass through bio-inspiration and micro-architecture. Nature Communications 2014 5:1 *5*, 1–9. https://doi.org/10.1038/ncomms4166.

9. Barthelat, F. (2015). Architectured materials in engineering and biology: fabrication, structure, mechanics and performance. https://doi.org/10.1179/1743280415Y.0000000008 *60*, 413–430.

10. Barthelat, F., Yin, Z., and Buehler, M.J. (2016). Structure and mechanics of interfaces in biological materials. Nature Reviews Materials 2016 1:4 *1*, 1–16. https://doi.org/10.1038/natrevmats.2016.7.

11. Achrai, B., Bar-On, B., and Wagner, H.D. (2014). Bending mechanics of the red-eared slider turtle carapace. J Mech Behav Biomed Mater *30*, 223–233. https://doi.org/10.1016/J.JMBBM.2013.09.009.

12. Barthelat, F., and Rabiei, R. (2011). Toughness amplification in natural composites. J Mech Phys Solids *59*, 829–840. https://doi.org/10.1016/J.JMPS.2011.01.001.

13. Espinosa, H.D., Rim, J.E., Barthelat, F., and Buehler, M.J. (2009). Merger of structure and material in nacre and bone – Perspectives on de novo biomimetic materials. Prog Mater Sci *54*, 1059–1100. https://doi.org/10.1016/J.PMATSCI.2009.05.001.

14. Chen, I.H., Yang, W., and Meyers, M.A. (2015). Leatherback sea turtle shell: A tough and flexible biological design. Acta Biomater *28*, 2–12. https://doi.org/10.1016/J.ACTBIO.2015.09.023.

15. Rivera, J., Hosseini, M.S., Restrepo, D., Murata, S., Vasile, D., Parkinson, D.Y., Barnard, H.S., Arakaki, A., Zavattieri, P., and Kisailus, D. (2020). Toughening mechanisms of the elytra of the diabolical ironclad beetle. Nature 2020 586:7830 *586*, 543–548. https://doi.org/10.1038/s41586-020-2813-8.

16. Jentzsch, M., Becker, S., Thielen, M., and Speck, T. (2022). Functional Anatomy, Impact Behavior and Energy Dissipation of the Peel of Citrus × limon: A Comparison of Citrus × limon and Citrus maxima. Plants 2022, Vol. 11, Page 991 *11*, 991. https://doi.org/10.3390/PLANTS11070991.

17. Thielen, M., Schmitt, C.N.Z., Eckert, S., Speck, T., and Seidel, R. (2013). Structure-function relationship of the foam-like pomelo peel (Citrus maxima)-an inspiration for the development of biomimetic damping materials with high energy dissipation. Bioinspir Biomim *8*. https://doi.org/10.1088/1748-3182/8/2/025001.

18. Lee, N., Horstemeyer, M.F., Rhee, H., Nabors, B., Liao, J., and Williams, L.N. (2014). Hierarchical multiscale structure–property relationships of the red-bellied woodpecker (Melanerpes carolinus) beak. J R Soc Interface *11*. https://doi.org/10.1098/RSIF.2014.0274.

19. Perricone, V., Sarmiento, E., Nguyen, A., Hughes, N.C., and Kisailus, D. (2024). The convergent design evolution of multiscale biomineralized structures in extinct and extant organisms. Communications Materials 2024 5:1 *5*, 1–18. https://doi.org/10.1038/s43246-024-00669-z.

20. Barthelat, F. (2010). Nacre from mollusk shells: a model for high-performance structural materials. Bioinspir Biomim *5*. https://doi.org/10.1088/1748-3182/5/3/035001.

21. Espinosa, H.D., Juster, A.L., Latourte, F.J., Loh, O.Y., Gregoire, D., and Zavattieri, P.D. (2011). Tablet-level origin of toughening in abalone shells and translation to synthetic composite materials. Nature Communications 2011 2:1 *2*, 1–9. https://doi.org/10.1038/ncomms1172.

22. Miura, T., Perlyn, C.A., Kinboshi, M., Ogihara, N., Kobayashi-Miura, M., Morriss-Kay, G.M., and Shiota, K. (2009). Mechanism of skull suture maintenance and interdigitation. J Anat *215*, 642–655. https://doi.org/10.1111/J.1469-7580.2009.01148.X.

23. Currey, J.D., Nash, A., and Bonfield, W. (1982). Calcified cuticle in the stomatopod smashing limb. J Mater Sci *17*, 1939–1944. https://doi.org/10.1007/BF00540410/METRICS.

24. Amorim, L., Santos, A., Nunes, J.P., and Viana, J.C. (2021). Bioinspired approaches for toughening of fibre reinforced polymer composites. Mater Des *199*, 109336. https://doi.org/10.1016/J.MATDES.2020.109336.

25. Dyskin, A. V., Estrin, Y., and Pasternak, E. (2019). Topological interlocking materials. Springer Series in Materials Science *282*, 23–49. https://doi.org/10.1007/978-3-030-11942-3_2/COVER.

26. Estrin, Y., Krishnamurthy, V.R., and Akleman, E. (2021). Design of architectured materials based on topological and geometrical interlocking. Journal of Materials Research and Technology *15*, 1165–1178. https://doi.org/10.1016/J.JMRT.2021.08.064.

27. Liu, K., Sun, R., and Daraio, C. (2022). Growth rules for irregular architected materials with programmable properties. Science (1979) *377*, 975–981. https://doi.org/10.1126/SCIENCE.ABN1459/SUPPL_FILE/SCIENCE.ABN1459_MOVIES_S1_TO_S3.ZIP.

28. Wang, R., Bian, Y., and Liu, K. (2025). Nonlinear mechanical properties of irregular architected materials. J Appl Mech, 1–20. https://doi.org/10.1115/1.4067570.

29. Jia, Y., Liu, K., and Zhang, X.S. (2024). Topology optimization of irregular multiscale structures with tunable responses using a virtual growth rule. Comput Methods Appl Mech Eng *425*, 116864. https://doi.org/10.1016/J.CMA.2024.116864.

30. Jia, Y., Liu, K., and Zhang, X.S. (2024). Unstructured growth of irregular architectures for optimized metastructures. J Mech Phys Solids *192*, 105787. https://doi.org/10.1016/J.JMPS.2024.105787.

31. Fox, C., Chen, K., Antonini, M., Magrini, T., and Daraio, C. (2024). Extracting Geometry and Topology of Orange Pericarps for the Design of Bioinspired Energy Absorbing Materials. Advanced Materials, 2405567. https://doi.org/10.1002/ADMA.202405567.

32. Thorpe, M.F. (1983). Continuous deformations in random networks. J Non Cryst Solids *57*, 355–370. https://doi.org/10.1016/0022-3093(83)90424-6.

33. Thorpe, M., and Duxbury, P. (2002). Rigidity theory and applications. Rigidity Theory and Applications. https://doi.org/10.1007/B115749.

34. Lu, T., Zhou, Z., Bordeenithikasem, P., Chung, N., Franco, D.F., Andrade, J.E., and Daraio, C. (2024). Role of friction and geometry in tuning the bending stiffness of topologically interlocking materials. Extreme Mech Lett *71*, 102212. https://doi.org/10.1016/J.EML.2024.102212.

35. Khandelwal, S., Siegmund, T., Cipra, R.J., and Bolton, J.S. (2015). Adaptive mechanical properties of topologically interlocking material systems. Smart Mater Struct *24*, 045037. https://doi.org/10.1088/0964-1726/24/4/045037.

36. Siegmund, T., Barthelat, F., Cipra, R., Habtour, E., and Riddick, J. (2016). Manufacture and Mechanics of Topologically Interlocked Material Assemblies. Appl Mech Rev *68*. https://doi.org/10.1115/1.4033967/370056.

37. Estrin, Y., Beygelzimer, Y., Kulagin, R., Gumbsch, P., Fratzl, P., Zhu, Y., and Hahn, H. (2021). Architecturing materials at mesoscale: some current trends. Mater Res Lett *9*, 399–421. https://doi.org/10.1080/21663831.2021.1961908.

38. Estrin, Y., Krishnamurthy, V.R., and Akleman, E. (2021). Design of architectured materials based on topological and geometrical interlocking. Journal of Materials Research and Technology *15*, 1165–1178. https://doi.org/10.1016/J.JMRT.2021.08.064.

39. Djumas, L., Simon, G.P., Estrin, Y., and Molotnikov, A. (2017). Deformation mechanics of non-planar topologically interlocked assemblies with structural hierarchy and varying geometry. Scientific Reports 2017 7:1 *7*, 1–11. https://doi.org/10.1038/s41598-017-12147-3.

40. Feldfogel, S., Karapiperis, K., Andrade, J., and Kammer, D.S. (2023). Scaling, saturation, and upper bounds in the failure of topologically interlocked structures. Int J Solids Struct *269*, 112228. https://doi.org/10.1016/J.IJSOLSTR.2023.112228.

41. Dyskin, A. V, Estrin, Y., Kanel-Belov, A.J., and Pasternak, E. Toughening by FragmentationÐ How Topology Helps**. https://doi.org/10.1002/1527-2648(200111)3:11.

42. Dyskin, A. V., Estrin, Y., Pasternak, E., Khor, H.C., and Kanel-Belov, A.J. (2003). Fracture Resistant Structures Based on Topological Interlocking with Non-planar Contacts. Adv Eng Mater *5*, 116–119. https://doi.org/10.1002/ADEM.200390016.

43. Mirkhalaf, M., Zhou, T., and Barthelat, F. (2018). Simultaneous improvements of strength and toughness in topologically interlocked ceramics. Proc Natl Acad Sci U S A *115*, 9128–9133. https://doi.org/10.1073/PNAS.1807272115/SUPPL_FILE/PNAS.1807272115.SAPP.PDF.

44. Sykes, M.F., Essam, ; J W, and Essam, J.W. (1964). Exact Critical Percolation Probabilities for Site and Bond Problems in Two Dimensions. J Math Phys *5*, 36. https://doi.org/10.1063/1.1704215.

45. Nash, R.J., and Li, Y. (2021). Experimental and numerical analysis of 3D printed suture joints under shearing load. Eng Fract Mech *253*, 107912. https://doi.org/10.1016/J.ENGFRACMECH.2021.107912.

46. Ershov, D., Phan, M.S., Pylvänäinen, J.W., Rigaud, S.U., Le Blanc, L., Charles-Orszag, A., Conway, J.R.W., Laine, R.F., Roy, N.H., Bonazzi, D., et al. (2022). TrackMate 7: integrating state-of-the-art segmentation algorithms into tracking pipelines. Nature Methods 2022 19:7 *19*, 829–832. https://doi.org/10.1038/s41592-022-01507-1.

47. Wegst, U.G.K., Bai, H., Saiz, E., Tomsia, A.P., and Ritchie, R.O. (2014). Bioinspired structural materials. Nature Materials 2014 14:1 *14*, 23–36. https://doi.org/10.1038/nmat4089.

48. Euler, L. (1758). Elementa doctrinae solidorum. Novi Commentarii academiae scientiarum Petropolitanae.

49. Dizon, J.R.C., Espera, A.H., Chen, Q., and Advincula, R.C. (2018). Mechanical characterization of 3D-printed polymers. Addit Manuf *20*, 44–67. https://doi.org/10.1016/J.ADDMA.2017.12.002.

50. Bell, D., and Siegmund, T. (2018). 3D-printed polymers exhibit a strength size effect. Addit Manuf *21*, 658–665. https://doi.org/10.1016/J.ADDMA.2018.04.013.

51. Slesarenko, V., and Rudykh, S. (2018). Towards mechanical characterization of soft digital materials for multimaterial 3D-printing. Int J Eng Sci *123*, 62–72. https://doi.org/10.1016/J.IJENGSCI.2017.11.011.

**Supplementary Information of:**

**Controllable Interlocking from Irregularity in Two-Phase Composites**

*Chelsea Fox[1], Kyrillos Bastawros[1], Tommaso Magrini[2*] and Chiara Daraio[1*]*

[1] Division of Engineering and Applied Science, California Institute of Technology, Pasadena, CA 91125, USA

[2] Department of Mechanical Engineering, Eindhoven University of Technology, 5600MB Eindhoven, The Netherlands

[*] Email: t.magrini@tue.nl, daraio@caltech.edu

**Figure S1**

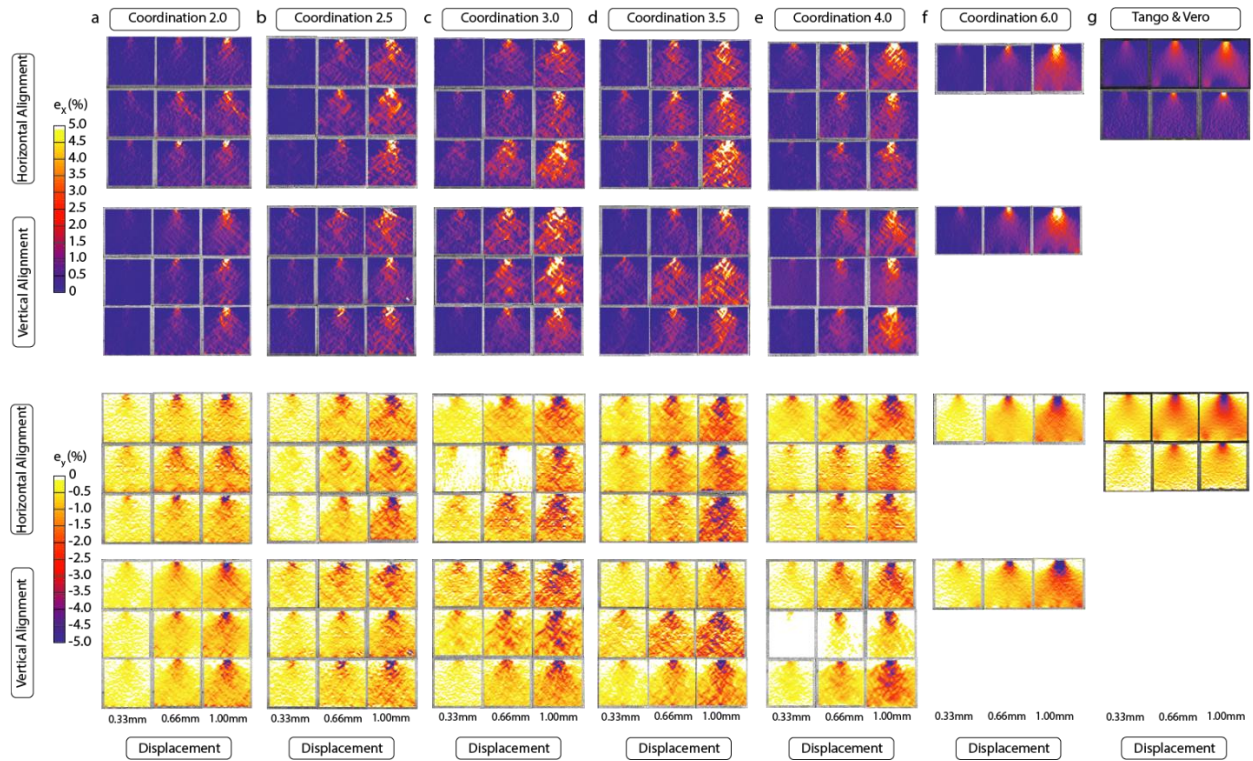

Figure S1. Cylindrical contact loading characterization. 2D DIC maps of all samples for $\varepsilon_x$ strain (above) and $\varepsilon_y$ strain (below).

**Figure S2**

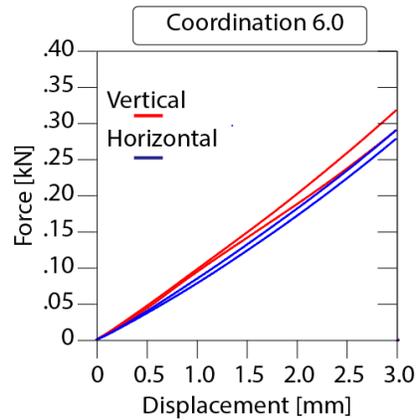

Figure S2. Force-displacement plots for coordination 6X, zoomed in to show variation from sample orientation.